# The 2025 June 01 Forbush Decrease measured over a range of primary cosmic ray energies

Roger Clay[1]

[1]*School of Physics, Chemistry and Earth Sciences, University of Adelaide, Adelaide, South Australia 5005, Australia*

roger.clay@adelaide.edu.au

**Abstract**

The 2025 June 01 Forbush Decrease in the terrestrial ground-level flux of cosmic ray secondaries was recorded by many cosmic ray systems. This was the deepest such decrease, from the quiescent value of the flux, which has been observed in the past two decades. It resulted from a complex series of solar events, none of which on its own reached the most extreme level. The extreme depth of this Decrease has enabled measurements of the flux reduction to be made, which would normally be severely limited by particle counting statistics. In particular, here we examine the Decrease phenomenon over a primary cosmic ray energy range which is rarely accessible, due to the low flux of high energy cosmic rays. This work considers data mainly from a muon telescope system which can respond to both unaccompanied muons and small cosmic ray air showers, providing data from GeV to mid-TeV energies, where the Forbush Decrease ceases to be statistically observable. This paper examines the depth of the flux Decrease, as a fraction below its quiescent value, over that primary energy range. The progressive development of internal time structure in the flux through the seven days of the phenomenon is also demonstrated.

**Keywords:** Forbush Decrease; June 2025; muon detection; high primary energy range

# 1. Introduction

Almost all cosmic rays which are recorded terrestrially originate outside our heliosphere. Very few are of solar origin. At primary particle energies below about 1 PeV ($10^{15}$ eV) the recorded cosmic rays are believed to come from sources within our galaxy [1,2,3]. Above those energies, it is believed that extra-galactic sources predominate. Charged cosmic rays (thus excluding photons and neutrinos) are nuclei which reach us after their paths are defined by magnetic fields, including those within the heliosphere. Heliospheric magnetic fields have little effect for the higher energy particles but the abundant lowest energy particles reach us preferentially along the spiral line directions defined by the heliospheric structure of the solar wind – the Parker spiral [4].

At the Earth, those lowest energy particles thus have a (slightly) preferred arrival direction defined by the heliospheric structure and the orbital motion of the Earth. As the Earth rotates over 24 hours, there may appear a daily oscillatory 'modulation' of the low energy cosmic ray flux with a broad maximum in the preferred cosmic ray arrival direction and a minimum 12 hours later. At times of a relatively quiet Sun, the flux maximum is found at mid-afternoon (local) times. The time variation of the primary cosmic ray flux at the lowest terrestrial energies is most often studied by monitoring the ground-level flux of neutrons using 'neutron monitors' but slightly higher energy primary cosmic rays can produce muons whose flux is detectable at the ground with muon telescopes. As the solar wind varies in speed and density, the amplitude of that daily modulation changes over a typical period of a few days with a characteristic amplitude up to ~1% of the overall flux. The Sun has longer-term periodicities which are again to be reflected in long-term changes of the overall ground-level flux of particles [5]. Short-term (instantaneous) solar events occur occasionally and are of particular interest. They result from interesting solar physics associated with the structure of the solar magnetic field, and can result in the formation of a plasma and magnetic field shock which provides a local laboratory for understanding more energetic events found throughout the Universe [6].

This "space weather" is routinely recorded in detail by instruments on satellites but those are necessarily small, and large-area ground-based detectors add data which can have better statistical properties. Also, space instruments have a limited ability to study a range of cosmic ray energies since the cosmic ray energy spectrum is steep, with an integral spectrum falling roughly as energy squared. In this paper, measurements are made of a solar event over a significantly larger range of primary cosmic ray energies. Such data are not available from other low energy cosmic ray telescopes.

From time to time, a solar event will result in the ejection of substantial plasma from a region of the Sun, such that it travels in a path which includes the Earth. That plasma mass (within the overall phenomenon known as a Coronal Mass Ejection (CME)) may travel with high speed and develop a dense plasma/magnetic shock front which is capable of partially blocking the terrestrial arrival of low energy galactic cosmic rays diffusing inwards through the heliosphere. The result is a sharp (hours or less) reduction in the flux of background (secondary) cosmic rays measured at the surface of the Earth with a recovery over several days. This is known as a Forbush Decrease (FD) [7,8]. In the context of this paper, Figure 3 of [8] (one of the earliest descriptions of this phenomenon with data recorded in 1937) exhibits striking similarities in structure (but at a lower amplitude) to data presented here.

A review describing general cosmic ray modulation and its observational techniques can be found in [9]. Measurements of the Forbush decrease phenomenon can be related to, and add to, our understanding of episodic heliospheric magnetic field enhancements following solar events. Incoming cosmic rays diffuse across those magnetic fields and their associated shock turbulence [10] with their observable flux changes reflecting the time varying structure of the local heliosphere.

Modulation in the form of Forbush Decreases, and less spectacular heliospheric events, are most commonly monitored at ground level by recording the flux of cosmic ray induced neutrons. Neutron monitors respond to low energy primary cosmic ray interactions in upper levels of the atmosphere in which cosmic ray protons (the most common species at these energies) interact with atmospheric particles to produce neutrons which then travel relatively easily to ground level. The incoming low energy protons are deflected by the terrestrial magnetic field and only those with energies (rigidities) above a (location-dependent) 'cut-off' can reach significant atmospheric levels and be recorded. This sets a primary energy threshold for neutron monitor data and such energies are commonly stated to be in the 10s of GeV range for those systems.

Cosmic ray particles interacting in the upper atmosphere (typically in the top 10%, the first 100 g.cm$^{-2}$) can also have particle-nucleus interactions producing secondary pions, provided that the initial cosmic ray energies are sufficiently high. Those pions may quickly decay to muons which travel down through the atmosphere relatively easily, predominantly only losing energy by ionisation at a rate of about 2 MeV /(g.cm$^{-2}$). For a vertical atmospheric thickness, this equates to about 2 GeV of energy loss and this, of course, increases with increasing zenith angle. The muons are produced in limited numbers in interactions at these energies and so, when arriving at the surface of the Earth, they are often detected as single particles and are referred to as 'unaccompanied' muons. That is, they are detected as single muon events. These detectable muons had a few tens of GeV energy at their source and the neutrons discussed above had a few GeV. Forbush Decreases measured with neutron monitors and with muon detectors (the latter possibly with directional beams at various zenith angles) can thus be studied through monitoring the flux of incoming <u>single</u> cosmic ray particles over a range of primary (GeV) energies [9,11].

If the primary cosmic ray particle has an appreciably higher energy (in the TeV range for the case in this paper), the number of produced pions will be greater. The neutral pions will produce gamma-rays which initiate electromagnetic (electron and photon) cascades and there will be larger accompanying numbers of muons. These, together with remnants of the primary particle, make up cosmic ray (extensive) air showers and can be detected by arrays of ground-based particle detectors. Such air showers differ from unaccompanied muons in their atmospheric absorption properties. The air showers decay with an absorption length of 150-200 g.cm$^{-2}$ but the muons are less absorbed, with absorption lengths 3-4 times greater [12]. This is reflected in their rate dependencies on atmospheric pressure, the barometric coefficients of the two phenomena. The muons have a rate attenuation of -0.13-0.15 % per millibar change of atmospheric pressure whilst air showers have barometric coefficients of about -0.5% /mb. This change is clear in data presented here and in long-term studies with the same equipment based on studies of the pressure coefficient as a function of signal amplitude, beginning at the signal level for a single muon [13]. At a signal amplitude level of several times that for a single muon, the coefficient becomes the value for showers (see [13] Figure 5.3.4). AIRES shower modeling presented in [13] indicates that small air showers, identified by a barometric coefficient of -0.5%/mb, such as studied here, have primary energies in the 1-10 TeV range.

This paper presents data for the exceptionally strong 2025 June 01 Forbush Decrease based on the operation of both a complex muon detector and small air shower system at the Buckland Park field station of the University of Adelaide, and a single one square meter muon detector 45 km to the south located at the University of Adelaide. The Buckland Park system responds to (a) single muons without directional constraint, plus (b) muons limited to the zenith and two larger zenith angles through detector coincidences. Additionally, (c) more complex coincidence combinations of the same Buckland Park detectors allow air shower measurements to be made into the TeV region of primary cosmic ray energies. Those data allow a more substantial energy range to be studied for the Forbush Decrease phenomenon than is possible with conventional monitors, which respond solely to single particles. The data presented here offer a unique view of a Fobush decrease through recording the phenomenon over a substantial range of primary cosmic ray energies.

**2. Materials and Methods**

The basic component of the radiation detector systems discussed in this paper is a one square meter plastic scintillator muon detector. Multiple such detectors make up the Buckland Park system discussed here. Each detector consist of a one square meter (50 mm thick) slab of plastic scintillator viewed 850 mm from above by a 120 mm photomultiplier tube. Large area plastic scintillators such as this are conventionally calibrated using the histogram of pulse heights from the untriggered detector. This is dominated by cosmic ray muon signals and exhibits a broad peak, normally well clear from noise. From this can be derived a signal amplitude equivalent to a single vertical muon. The detectors here have signal discriminator thresholds set to respond to single (and above) vertical muons. The resulting muon rate is close to one particle per square centimeter per minute, or about 160 Hz for a one square meter scintillator. If counts are accumulated in 15 minute intervals (as in this work), each detector then has a mean count of about 150,000 and a Poisson uncertainty of a little under 0.3%. There is one of these scintillator units at Adelaide in a room below ~100 g.cm$^{-2}$ of building material, and there is a complex system of these below a light roof at the Buckland Park site.

At Buckland Park (sea-level, 34.5º South, 138.5º East, geomagnetic cut-off 4 GV), there are one square meter scintillator units arranged in a square as shown in plan in Figure 1. There are four units as shown in Figure 1. Each of the scintillators shown in the Figure is directly above (0.94 m) an identical detector. Those further four units complete an eight scintillator muon telescope. These individually insulated detectors are housed in a thermally insulated light-tight room.

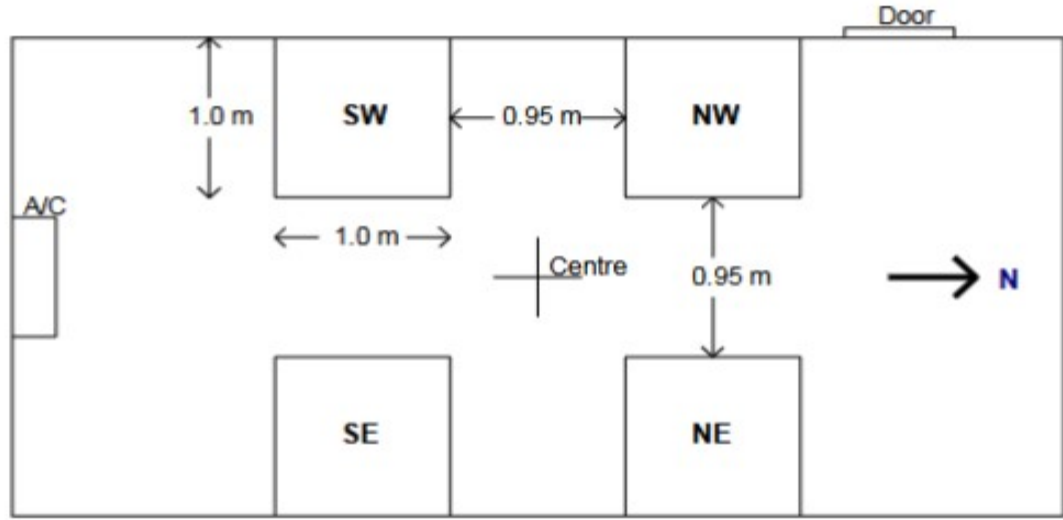

**Figure 1**. Plan of the Buckland Park muon telescope. The internal one square meter squares represent the locations of four scintillator units which are vertically (directly) above another similar set of four.

The signals from the scintillators are taken to discriminators which produce 1µs pulses when triggered on signals corresponding to single muons. Those pulses are received by a field programmable gate array (FPGA) which applies logic to the eight input channels, producing a number of individual detector and coincidence signals. These are counted and recorded in 15 minute intervals. A single muon passing through one of the upper scintillators and <u>also</u> a lower scintillator produces a coincidence signal which can be used to define (roughly) an arrival direction. The different arrival zenith angles lead to mean energy difference estimates between 'vertical' and 'diagonal' primary particles. The data acquired by the FPGA are:

1. Eight individual detector counts.

2. A <u>total</u> of the eight individual counts

3. The sum of coincidences between of the four pairs of vertically located detectors.

4. The sum of coincidences between each of north, south, east, and west pointing diagonals (for instance, coincidences between upper detectors SE and NE and lower corresponding detectors SW and NW produce an easterly muon beam). These north, south, east and west beams are called here "Short" diagonals.

5. The sum of coincidences for four "Long" diagonal beams. For instance the upper SE detector and the lower NW detector (south east, south west etc. beams).

The coincidences in items 4 and 5 could accidentally be the result of independent particles in small showers rather than a single diagonally travelling muon. This is checked by counting all diagonal events (items 4 and 5) which had an additional detector triggered in coincidence. This would only be possible if more than one particle had taken part in producing the coincidence. For instance the simultaneous arrival of two independent shower particles (characteristic of a small air shower) could trigger a diagonal coincidence but one of those, at least, may well trigger a further detector, above or below the ones in the original coincidence. We call these events, which are <u>not</u> produced by single unaccompanied muons, "vetoes" and subtract their rates from the original diagonal rates, so that we have correct unaccompanied muon diagonal rates. We also, then, have a rate measure for what are likely to be small air showers, identified by the simultaneous arrival of multiple particles. We therefore have data corresponding to small cosmic ray showers through event coincidences which are not possible with single muons alone:

6. Veto rates, corresponding to the arrival of small air showers.

Small air showers can trigger more than just two diagonal detectors. These give the rates in item 6, but small showers can also be identified by other forms of multiple coincidences which are not possible for single unaccompanied muons.

We sum rates of:

7. Two-fold coincidences in either top or bottom sets of detectors,

8. Three-fold coincidences in either top or bottom sets of detectors,

9. Four-fold coincidences in either top and bottom sets of detectors plus

10. Eight-fold coincidence rate for all scintillator detectors triggering in coincidence. Whilst not used here, we note that these eight-fold coincidences are GPS time stamped for other studies.

The Adelaide single one-square-meter muon detector also triggers on single muons with a rate similar to that for the individual Buckland Park detectors. It is small but it can be used to confirm any significant time structure found in counts 1 and 2 above.

As noted earlier, the muon data need correction for variations in atmospheric pressure with coefficients of about -0.13-0.15 % per millibar change of atmospheric pressure for single muon detection and about -0.5% /mb for air showers. It was not obvious that those values should necessarily apply to the recordings within a Forbush decrease, for which the incoming cosmic ray energy spectrum is unknown, and so their values were checked for each dataset. The corrections can be determined rather clearly and the uncertainties are ~0.02%/mb and ~0.1%/mb respectively for the single muon and air shower cases. The results are, in fact, consistent with expectation from measurements at undisturbed times.

**3. Results**

The 2025 June 01 Forbush Decrease was recorded in many of the channels described above. The Buckland Park total count summed for all eight detectors had the highest rate and hence smallest Poisson uncertainties (Figure 2a,b.). It can be seen that here is a clear rapid fall in rate, followed by a recovery phase, and a variation within it, which appears to have a roughly diurnal period (see below). It is noteworthy that successive records in the 15 minute time sequence do not have variations which exceed the 0.1%Poisson level limitation for ~ 1 million counts in an interval. Thus, there are no clear short-term high frequencies at the 900s time-scale level.

a.

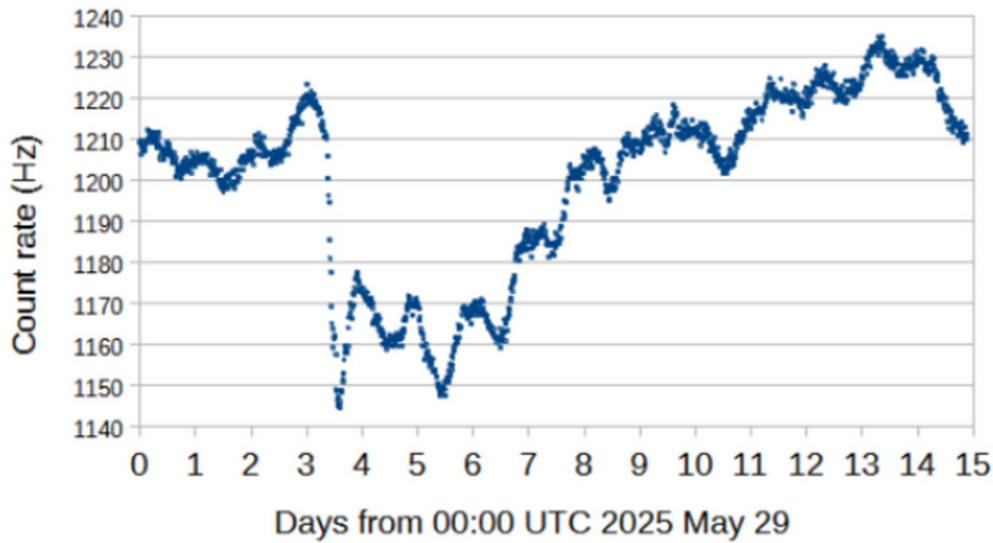

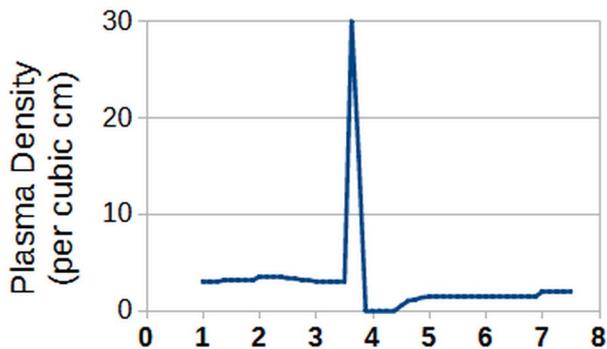

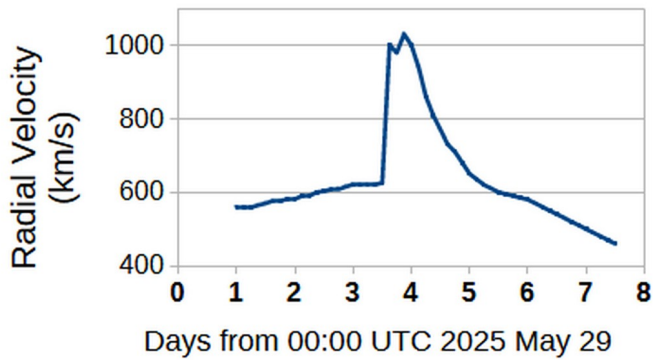

b.

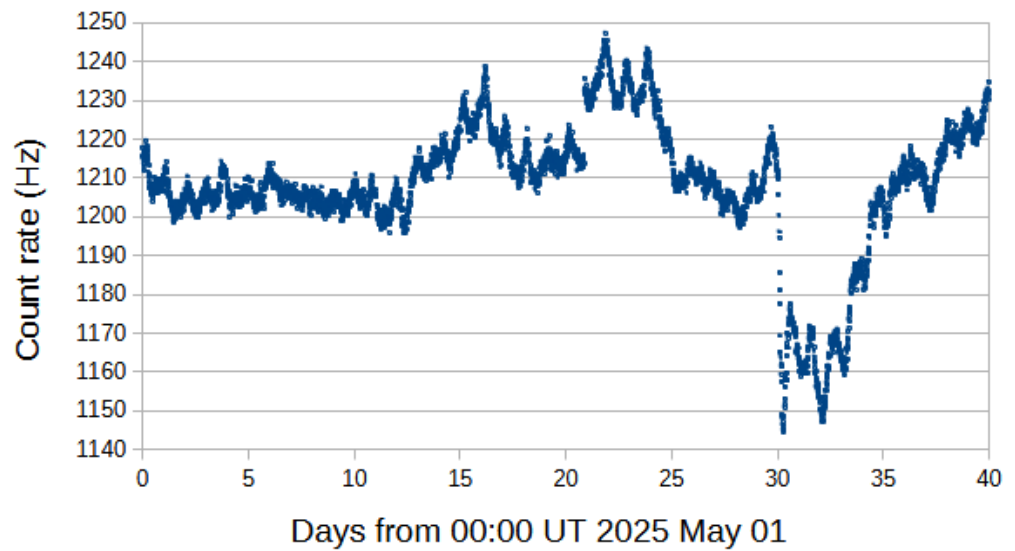

**Figure 2**. Buckland Park total rate for the sum of all eight detectors, recorded at 900 s intervals.  **a.** Muon data covering the Forbush decrease. Also, space plasma, solar wind, density and radial velocity data redrawn with 3h resolution from the NOAA Space Weather Prediction Centre are added to show spacecraft data from the corresponding period  (https://www.swpc.noaa.gov/)  **b.** Extending the time range, beginning at the start of May, to indicate levels of other rate variations.

As a check that the systems work correctly, the Adelaide data (Figure 3.) located 45 km to the South can be also examined in comparison with Figure 2a. This has only one eighth of the detector area compared with Buckland Park and it is located below building material acting as a modest absorber. The agreement is reassuring although, clearly, there is some small non-identical structure.

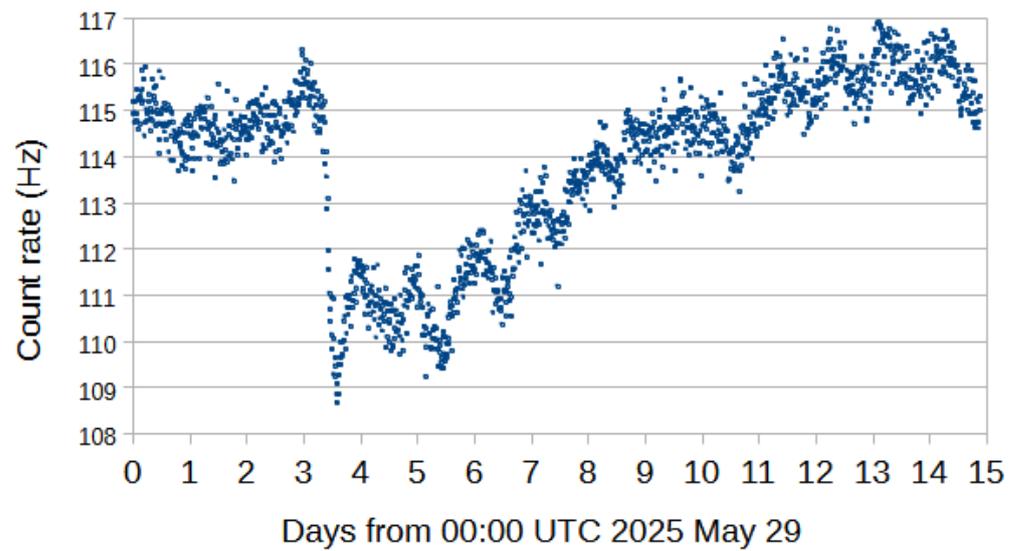

**Figure 3**. Rates recorded at 900s intervals by the Adelaide single square meter detector located 45 km south of Buckland Park.

Muons which are recorded by simple detectors have a very broad zenith angle distribution and a more precise vertical angular field of view can be obtained with the coincidence arrangement listed as '3.' above. Those data are shown in Figure 4.

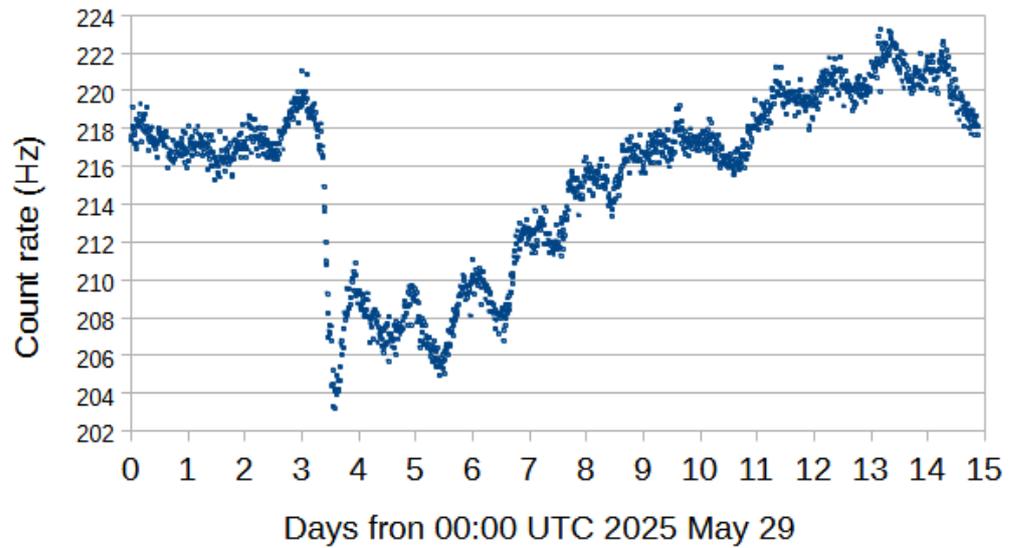

**Figure 4.** Data with four channels of coincidences determining a 'vertical' muon beam at Buckland Park. Rates were recorded at 900 s intervals.

The initial downward edge of the Forbush Decrease in Figure 4 appears rather smooth. Detail from Figure 4 is shown in Figure 5 which shows that, with the available time resolution of 900s, the decrease is indeed generally smooth but there is a 1% rate excess above the extrapolated fall towards the minimum in the Decrease, at around 12:15 UTC 01 June 2025, which lasts about 90 minutes.[1]

---

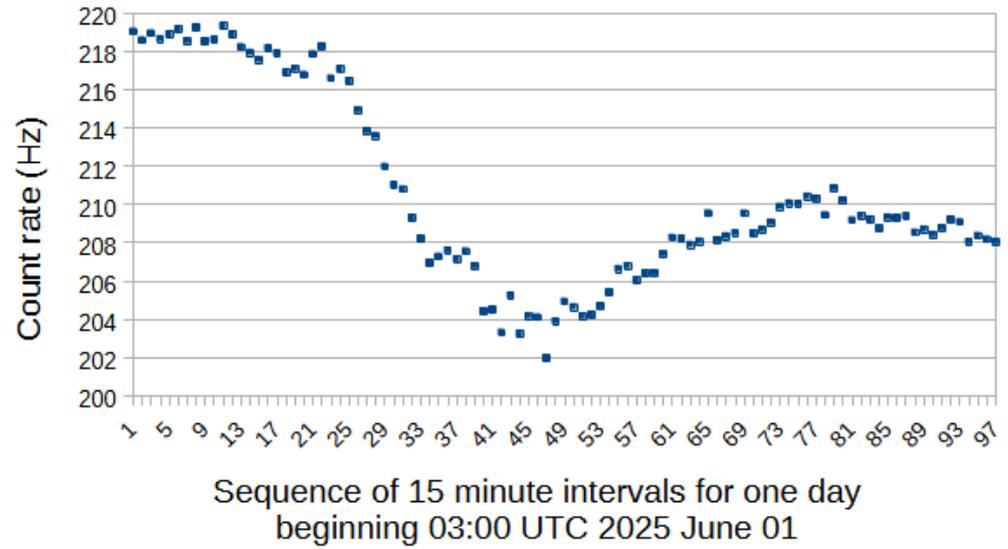

**Figure 5.** Detail from Figure 4 showing the rate variation over the first day of the Forbush decrease.

A coincidence arrangement (number 4. above) counts muons (with 'vetoes' subtracted) at larger zenith angles, with rather higher primary energies since they have passed through an increased atmospheric path. A pressure coefficient of 0.15 %/mb is still found to apply. Data for these counts are shown in Figure 6. The count rate is lower but the structure of the Decrease is still visible (these data were averaged over a moving window of five time periods).

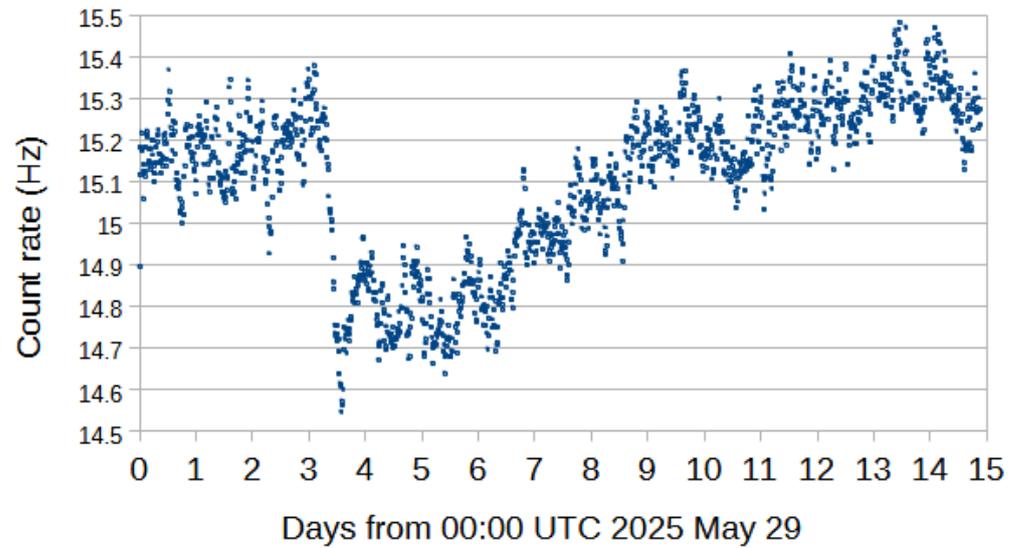

**Figure 6**. Counting data summed for all 'short' diagonals with 5 interval smoothing (characteristic zenith angle 62°). Data are presented at 900 s intervals

Data described above have been for the case of single muons. As previously noted, coincidences can be found which correspond to small air showers through the detection of the coincident arrival of multiple particles. The simplest of these is '7' above, giving data for two-fold coincidences, summed for either the top or bottom detector layer, at Buckland Park. Figure 7 shows such data with smoothing using a 5 time-period moving average. The same features are still present with, possibly, a stronger diurnal variation relative to the quiescent level in the recovery phase.

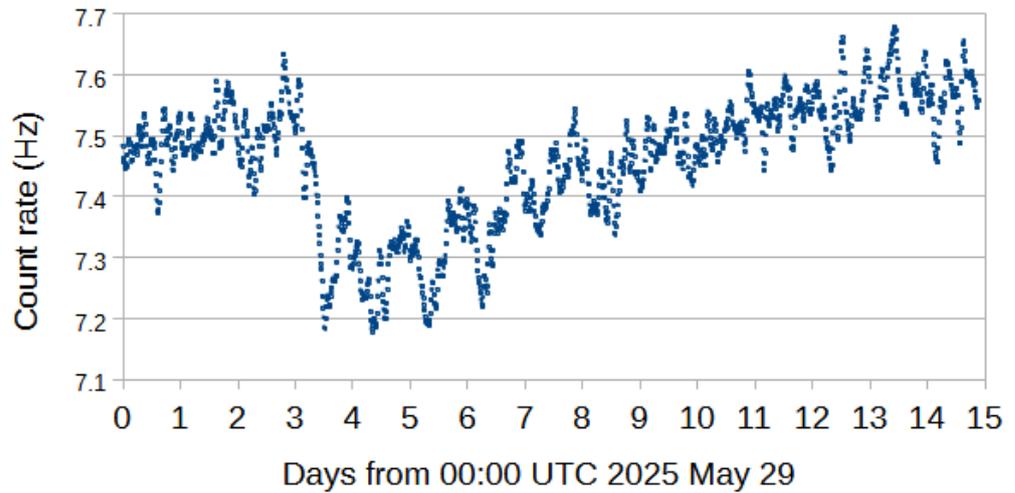

**Figure 7.** Two-fold coincidence rates at Buckland Park including for both top and bottom groups of four detectors. Rates are at 900 s intervals.

It was noted above that 'veto' channels correspond to air shower contamination of the diagonals datasets and so represent much higher primary cosmic ray energies than for the single muon detectors or neutron monitors. Those air showers are produced by primary cosmic rays in the TeV energy region, at ~10-100x higher energies than those associated with single muons. Figure 8 shows the summed vetoes for the short diagonal channels (smoothing with a five interval moving average). The statistical limits are worse here but the Forbush Decrease is still clear.

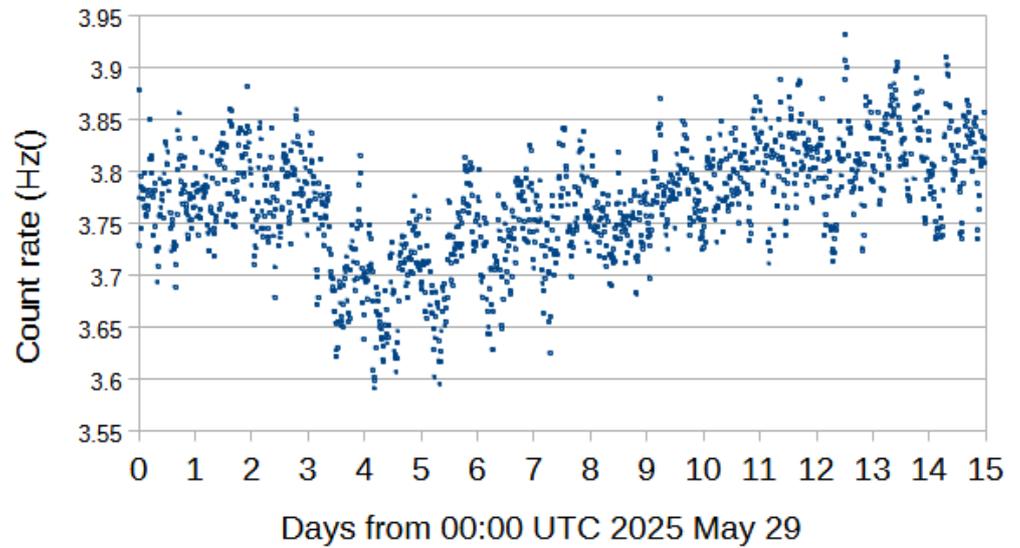

**Figure 8**. Summed veto channels for the 'short' diagonal dataset at Buckland Park. Rates data are at 900 s intervals.

The recorded dataset corresponding to the highest primary cosmic ray energies in which there is evidence for a Forbush Decrease is the three-fold channel (summing data both for top and bottom detector layers) and is shown in Figure 9 (eleven interval smoothing). There is no obvious evidence for a Forbush Decrease in four-fold coincidences at even higher primary cosmic ray energies. These have a much lower count rate and resulting low statistics.

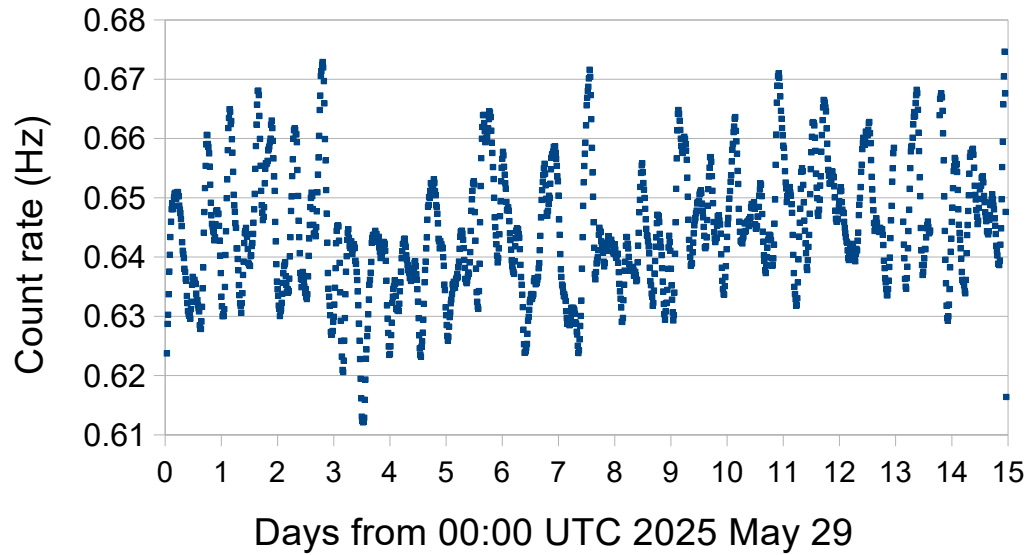

**Figure 9**. Three-fold coincidence data at 900 s intervals, summed for the upper and lower detector levels, recorded at Buckland Park. These data have been smoothed with an eleven interval moving average.

Figures 2 to 8 show a significant variation within the Forbush Decrease period, which at first sight may appear close to diurnal. A diurnal variation is often found in periods when the Sun is quiescent and the solar wind is stable, such that galactic cosmic rays follow a well defined inward path and have an anisotropic arrival direction distribution with a resulting 24 h cycle (see, for instance, days 21-24 of Figure 2b). Uncertainties when locating peaks and troughs (~1.5 h) due to statistics and some complex structure in the dataset (particularly at the minimum close to entry 05:00:00 03/06/2025 in Figure 2a) suggest that care in identifying the timing of those features is required. These locations have been identified either by recording the extrema in the data, where thay are clear, or fitting a local polynomial and locating its relevant maximum or minimum location. However, there is clear evidence that those variations are found to change in a systematic way with time, suggesting the existence of complex structure within the Decrease phenomenon. This may be associated with the idea of structure within the surrounding plasma magnetic field [14]. Figure 10 shows how the times (hours UTC) of the peaks and troughs in the muon data (Buckland Park total, Figure 2a) change over the first days of the Forbush Decrease whilst they still are clearly visible. There is a progressive change in their daily timing as the Decrease progresses. We note that at relatively quiet solar periods, the times of the peaks with this system are ~ 04:15 UTC (minima ~17:00 UTC), 13:45 local time.

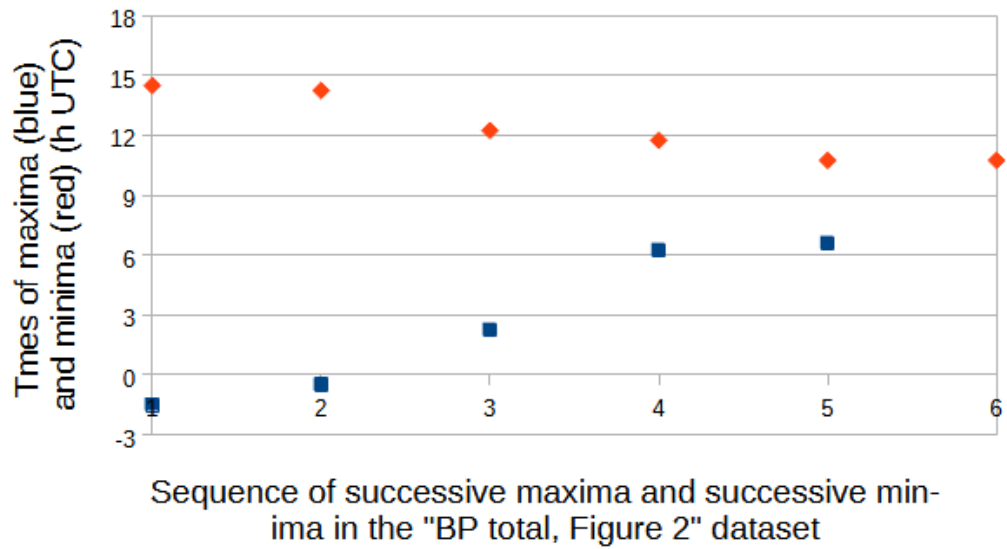

**Figure 10**. Times (h UTC – add 9.5 h for local times) of six successive peaks (blue) and five troughs (minima) (red) between them in the time sequence for Buckland Park total counts. The first trough is the initial one of the sharp decrease, followed by the first peak. The peaks and troughs are identified by sequential integers rather than their relative times (which progresively change in phase).

## 4. Discussion

The 2025 June 01 Forbush decrease was in response to a complex combination of solar events such that no single source component of the event reached nominally exceptional levels but the sum produced the largest Forbush decrease in the low energy ground-based cosmic ray flux for two decades. This extreme phenomenon was observed by many neutron and muon systems [11,15]. The event has recently been discussed in some detail by Chilingarian *et al.* [16]. It began most clearly with a coronal hole stream on May 29 2025 which was followed by four M-class flares over May 30-31, the most powerful being at a level of M8.2. Whilst none of the events were in the X-class, their sum produced an exceptional Forbush Decrease. The data presented by Chilingarian *et* *al.* are from a different terrestrial latitue and their data provide a different view of the Decrease to that presented here. The Decrease was clear at all recording sites [15], but the details within the Decrease, such as any oscillations, are local phenomena dependent on the particular view of the CME plasma. An understanding of such details will require much further work, but is underway. Figures 2-4 show detail that in the overall Decrease phenomenon, there is the appearance of a 'pre-decrease' followed by a 'pre-increase' in the period <u>before</u> the major flux reduction. Lingri et al. [17] have identified structure such as this as the result of the forward passage of the solar wind shock associated with the CME, followed by the resulting reflection of cosmic rays.

Compared to most other Forbush Decreases, the large fractional drop in flux of this Decrease means that the Poisson statistical limitations of measurements corresponding to high energy primary particles are less pronounced when compared to the drop in overall counts in the Decrease. It is of interest to determine what is the primary (proton) energy at which the phenomenon ceases to be observable. This will partly be determined by the steep primary particle energy spectrum but, also, it will relate to heliospheric magnetic field strengths and structure, including the overall dimensions of the CME plasma. The structure observed in a Forbush Decrease will be unique to the site of each terrestrial particle telescope since each has a unique viewing direction and related observation timing whilst the Earth rotates. It is a ongoing aim of current work to try to understand the relationship between Buckland Park 'diagonal' data and data from other directional telescopes, in order to obtain more complete full sky data in directions away from the ecliptic sites of spacecraft measurements. When complete, this will be reported elsewhere,

It is not easy to identify values of the mean primary energies corresponding to the particular recording channels described in this paper, even though muon generation and detection is generally understood, since the relevant spectrum of the primary particles is not known when under the influence of the Forbush decrease. However, broad statements of any energy dependence can be made based on conventional understandings of secondary cosmic ray properties, through modeling and experiment, of showers at related primary energies (see[13] [18]). The data from the various recording channels discussed above can be put in a sequence of increasing primary energies but not at well-defined energies. At sea-level, the muon energy spectrum is in the 10-100 GeV range for vertical muons. The primary particle energies for these particles are substantially higher than those values and in the 100s of GeV range for the large zenith angle 'diagonal' events. The primary energies for small showers (multiple ground-level particles) are in the multiple TeV range.

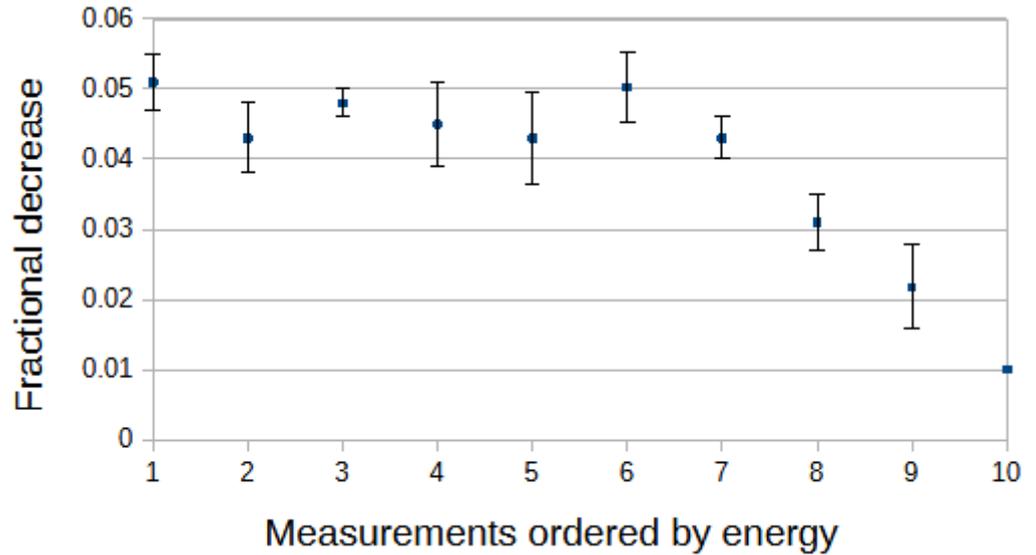

**Figure 11**. Sequence of progressively higher energy measurements of the fractional decrease found in the 01 June 2025 Forbush Decrease event. In order, these data are for: 1. Buckland Park vertical, 2. Adelaide data, 3. Buckland Park total detectors summed, 4. Short diagonals, 5. Long diagonals, 6. Short diagonal vetoes., 7. Two-folds, 8. Long diagonal vetoes, 9. Three-fold coincidences, 10. Four-fold coincidences (upper limit).

Figure 11 shows a sequence of measurements reported here of the fractional magnitude of the Forbush decrease expressed in sequential order of increasing primary particle (assumed protons) energies.

In determining these fractional decreases, note that the initial fall of the count rate in Figures 2-6 appears to 'overshoot'. When data are smoothed, as in some of the data presented, the depth of the overshoot will depend on the degree of smoothing. Due to this effect of smoothing on the initial narrow minimum, the data in Figure 11 correspond to decreases which ignore that initial sharp overshoot. For this reason, some data for lower energy entries in Figure 11 is smaller in amplitude than would normally be quoted. However, the single muon data are not smoothed and, for instance, the first channel in the figure would otherwise be a fractional decrease of 0.075 which would be appropriate for comparison with neutron monitor data which show typical decreases of ~0.12. The uncertainties shown in Figure 11 are estimated from a combination of the variations in the amplitudes of the quiescent flux before the Forbush Decrease and the statistical spread in data at the minimum of the Decrease.

The first five entries in Figure 11 use pressure coefficients ~0.15 %/mb and the remaining five correspond to small showers and use ~0.5 %/mb. These coefficients are those which would be expected for muons and air showers respectively but were derived here from the Forbush Decrease data themselves.

The break at about entry 7 in Figure 11 (two-fold triggers at Buckland Park) corresponds to primary proton energies ~3 TeV and entry 9 (the highest energy with an observable Forbush Decrease) corresponds to ~5 TeV protons. For a heliospheric magnetic field of ~1 nT, 3 TeV protons would have a radius of gyration somewhat larger than 1 AU but, at the time of the Forbush decrease, the magnetic field could well be ~100 nT over a more limited region, resulting in a radius of gyration appreciably less than 1 AU. Thus, the break energy seems to be physically plausible for a strong Forbush decrease such as this and may define physical parameters of the plasma cloud.

**5. Conclusions**

The 2025 June 01 Forbush Decrease provided an opportunity to observe the primary cosmic ray energy range which responds to the Forbush Decrease phenomenon. Forbush Decrease data have not previously been recorded over such a wide primary energy range but the measurements were possible here due to the exceptional nature of the event. Measurements have been presented here of the time and energy structure of the Decrease, over a period of up to ten days, and at energies from GeV up to the multi-TeV range, in datasets with counting rates covering a range of over a factor of 1000. That time and energy structure reflects the overall local heliospheric plasma and magnetic field structure associated with the event. Over seven days from start of the main decrease, that original deep decrease relaxed to very small values below quiescence (relaxing to half the original depth in three days) whist, over that time, some internal structure in the muon and air shower data (but not neutron monitor data at the lowest cosmic rays energies) changed appreciably in terms of the diurnal timing of clear peaks and troughs. This probably reflects changing internal structure within the plasma cloud.

## 6. Acknowledgements

## 7. Funding

This research received no external funding.

## 8. Data Availability Statement

Data used in this analysis are available on request from the author.

## 9. References

1. Linsley, J.;Scarsi, L.; Rossi, B. Extremely Energetic Cosmic-Ray Event. *Phys. Rev. Lett.* **1961,** *6,* 485-487.

2. The Pierre Auger Collaboration. Observations of a large-scale anisotropy in the arrival directions of cosmic rays above 8 x $10^{18}$ eV. *Science* **2017,** *357,* 1266-1270.

3. Beatty, J.J.; Matthews, J.; Wakely, S.P. 30. Cosmic Rays. *Particle Data Group* **2021,** *https://pdg.lbl.gov/2021/reviews/rpp2021-rev-cosmic-rays.pdf* (Accesssed 2025 September 15)

4. Parker, E.N. Dynamics of the Interplanetary Gas and Magnetic Fields. *Ap.J.* **1958,** *128***,** 664-676.

5. The Pierre Auger Collaboration. Scaler rates from the Pierre Auger Observatory: a new proxy of solar activity. *Ap.J.* **2025,** *987,* 41-53.

6. Protheroe, R.J.; Clay, R.W. Ultra High Energy Cosmci Rays. *Pub. Astron. Soc. Aust.* **2004,** *21,* 1-22.

7. Forbush, S.E. On the Effects in Cosmic-Ray Intensity Observed During the Recent Magnetic Storm. *Phys. Rev.* **1937***, 51,* 1108-1109.

8. Forbush, S.E. On cosmic ray effects associated with magnetic storms. *Terr. Mag.* **1938***, 43,* 205-2183.

9.  Duldig, M.L. Australian Cosmic Ray Modulation Research. *Pub. Astron. Soc. Aust.* **2001**, *18*, 12-40.

10.  Arunbabu, K.P.; Antia, H.M.; Dugad, S.R.; Gupta, S.K.; Hayashi, Y.; Kawakami, S.; Mohanty, P.K.; Oshima, A.; Subramanian, P.  How are Forbush decreases related to interplanetary magnetic field enhancements? *A&A* **2015**, *580*, A41-54.

11.  Zhang, J.L.; Tan, Y.H.; Wang, H.; Lu, H.; Meng, X.C.; Muraki, Y. The Yangbajing Muon-Neutron Telescope  *NIM A* **2010**, 1030-1034.

12.  Clay, R,; Gerhardy, P.R. Sea level measurements of properties of cosmic ray showers with sizes ranging from 200,000 to 10 million particles. *Aust. J. Phys.***1982,**

13.   Barber, K.B. Pierre Auger Observatory: Low Energy Cosmic Ray Anisotropies, PhD, University of Adelaide, Australia, **July 2014**  http://hdl.handle.net/2440/85987

14.  Laitinen, T.; Dalla, S. Access of Energetic Particles to a Magnetic Flux Rope from External Magnetic Field Lines. *Ap.J.* **2021**, *906.9 (10pp)*

15. Vanke, V.G. Quick Look Cosmic Rays (Moscow Neutron Monitor) and Important Links *IZMIRAN* **2025,** http://cosrays.izmiran.ru/   *(*Accesssed 2025 September 15*)*

16.   Chilingarian, A.: Karapetyan, T.; Sargsyan, B. The largest Forbush Decrease in 20 years: Preliminary analysis of SEVAN network observations. arXiv:2506.17917v1 (astro-ph.SR) **2025.**

17.  Lingri, D.; Mavromichalaki, H.; Belov, A.;Eroshenko, E.; Yanke,V.;Abunin,A.;Abunina,M. Solar Activity Parameters and Associated Forbush Decreases During the Minimum Between Cycles 23 and 24 and the Ascending Phase of Cycle 24 *Solar Physics* **2016,** *291,*1025-1042.

18. Clay, R. Measurements of Decoherence in Small Sea-Level Extensive Air Showers. *Universe* **2024,** *10* 308-321